\documentclass[aps,prb,twocolumn,superscriptaddress,longbibliography,floatfix]{revtex4-1}

\usepackage{amsmath}
\usepackage{graphicx}
\usepackage[colorlinks,linkcolor=blue,citecolor=blue,urlcolor=blue]{hyperref}
\usepackage[dvipsnames]{xcolor}
\usepackage{stix}
\usepackage[normalem]{ulem}
\begin{document}

\newcommand{\kk}{\mathbf{k}}
\newcommand{\qq}{\mathbf{q}}
\newcommand{\iwn}{i\omega_n}
\newcommand{\iwm}{i\omega_m}
\newcommand{\imG}{\mathrm{Im} G^R(\kk, z)}
\newcommand{\ivn}{i\nu_n}

\preprint{APS/123-QED}

\title{Spectral properties and enhanced superconductivity in renormalized Migdal-Eliashberg theory}% 

\author{Benjamin Nosarzewski}
\affiliation{%
Department of Physics, Stanford University, Stanford, California 94305, USA
}%
\affiliation{%
Stanford Institute for Materials and Energy Sciences, SLAC National Accelerator Laboratory and Stanford University, 2575 Sand Hill Road, Menlo Park, California 94025, USA
}%
\author{Michael Sch\"{u}ler}%
\affiliation{%
Stanford Institute for Materials and Energy Sciences, SLAC National Accelerator Laboratory and Stanford University, 2575 Sand Hill Road, Menlo Park, California 94025, USA
}%
\author{Thomas P. Devereaux}
\affiliation{%
Stanford Institute for Materials and Energy Sciences, SLAC National Accelerator Laboratory and Stanford University, 2575 Sand Hill Road, Menlo Park, California 94025, USA
}%
\affiliation{%
Department of Materials Science and Engineering, Stanford University, Stanford, California 94305, USA
}%
\affiliation{%
Geballe Laboratory for Advanced Materials, Stanford University, Stanford, California 94305, USA
}%

\date{\today}% It is always \today, today,

\begin{abstract}
 Migdal-Eliashberg theory describes the properties of the normal and superconducting states of electron-phonon mediated superconductors based on a perturbative treatment of the electron-phonon interactions. It is necessary to include both electron and phonon self-energies self-consistently in Migdal-Eliashberg theory in order to match numerically exact results from determinantal quantum Monte Carlo in the adiabatic limit. In this work we provide a method to obtain the real-axis solutions of the Migdal-Eliashberg equations with electron and phonon self-energies calculated self-consistently. Our method avoids the typical challenge of computing cumbersome singular integrals on the real axis and is numerically stable and exhibits fast convergence. Analyzing the resulting real-frequency spectra and self-energies of the two-dimensional Holstein model, we find that self-consistently including the lowest-order correction to the phonon self-energy significantly affects the solution of the Migdal-Eliashberg equations. The calculation captures the broadness of the spectral function, renormalization of the phonon dispersion, enhanced effective electron-phonon coupling strength, minimal increase in the electron effective mass, and the enhancement of superconductivity which manifests as a superconducting ground state despite strong competition with charge-density-wave order. We discuss surprising differences in two common definitions of the electron-phonon coupling strength derived from the electron mass and the density of states, quantities which are accessible through experiments such as angle-resolved photoemission spectroscopy and electron tunneling. An approximate upper bound on $2\Delta / T_c$ for conventional superconductors mediated by retarded electron-phonon interactions is proposed.

\end{abstract}

\maketitle

%\tableofcontents

\section{\label{sec:level1}Introduction}

The first microscopic theory of superconductivity was developed by Bardeen, Cooper, and Schrieffer (BCS theory) in 1957 \cite{BCS}. BCS theory assumes an instantaneous and non-local interaction between any pair of electrons within the Debye energy of the Fermi surface and this theory turns out to be an inadequate description for superconductors where the electron-phonon (el-ph) interaction is strong. The subsequent theory of superconductivity developed by Migdal and Eliashberg is designed to work at stronger el-ph coupling and considers a more realistic el-ph interaction which is retarded in time and successfully captures the frequency and momentum dependence el-ph coupling induced lifetimes and mass enhancement of electronic quasiparticle states near the Fermi level \cite{Ummarino, Migdal, Eliashberg}. 

There are different versions of Migdal-Eliashberg (ME) theory which differ in their level of approximation depending on the choice of whether to solve the ME equations self-consistently and whether to include the renormalization of the phonon propagator. Our interest here is in the self-consistent versions of ME theory and we will refer to the version which accounts for renormalization of the phonon propagator as renormalized Migdal-Eliashberg (RME) theory and the version where the phonon propagator remains bare as unrenormalized Migdal-Eliashberg (UME) theory. Previous work demonstrates remarkably good quantitative agreement between RME theory and numerically exact determinantal quantum Monte Carlo (DQMC) simulations on clusters of the same size in the limit where the dimensionless el-ph coupling is sufficiently weak and the phonon energy scale is small compared to the electronic bandwidth \cite{MarsiglioMC, Ilyame}. In particular, RME theory accurately captures the properties of the superconducting state of the system up to a critical value of the electron-phonon coupling strength beyond which ME theory breaks down due to lattice instabilities such as the formation of bipolarons or charge-density wave order \cite{Ilyame, ilyapg}. The self-consistent version of ME without renormalization of the phonon propagator is less accurate \cite{MarsiglioMC}. 

ME theory is commonly used to phenomenologically describe properties of the normal and superconducting states arising from el-ph interactions across many different materials. ME theory has been successfully applied to understand the superconducting transition temperature in $\mathrm{LaH}_{10}$ under high pressure \cite{drozdov, elatresh}, reproduce the single-particle tunneling density of states in lead \cite{Scalapino}, and provide theoretical support in the analysis of electronic band renormalizations seen in angle-resolved photo-emission (ARPES) experiments in strongly correlated systems such as the cuprates \cite{cuk}. In some applications, self-consistent ME solutions are needed as in the case of reproducing replica bands arising from a forward scattering electron-phonon interaction at the interface of monolayer FeSe on an $\mathrm{SrTiO_3}$ substrate or graphene/h-BN heterostructures \cite{JJ, yao}, and reproducing the phonon peak positions arising from multi-phonon scattering processes observed in high-resolution electron-tunneling spectroscopy on Nb-doped $\mathrm{SrTiO_3}$ \cite{harold}. With advances in inelastic x-ray scattering and neutron scattering, experiments are able to more accurately measure the dispersion and linewidths of phonons \cite{neutron, ixs}. Real-frequency solutions of the RME equations could prove to be a useful tool for understanding the phenomenology of phonon spectral functions.

The ME equations on the imaginary axis can be naturally solved via numerical methods thanks to the fact that the Matsubara frequencies are discretely spaced on the imaginary axis requiring simple summations and no integrations of functions with singularities. Furthermore, the imaginary axis solutions are numerically stable which is especially important when attempting to perform self-consistent calculations in challenging regimes such as at low temperature or moderately strong interaction strength. However, obtaining the solution of ME theory on the real axis in principle provides access to any property of the normal or superconducting state of the system and allows for direct comparison with experimentally accessible quantities such as the electron mass enhancement or the superconducting gap size. Unfortunately, the usual form of the ME equations on the real axis are much more cumbersome to solve numerically because they require evaluating principal-value integrals on every iteration and suffer from slower convergence. 

To avoid the difficulties of solving the ME equations on the real axis, it is common to use Pad\'e approximants to analytically continue the imaginary axis solution to the real axis or even to avoid analytic continuation entirely by using imaginary axis proxies such as correlators evaluated at $\tau=\beta/2$ \cite{vidberg, MarsiglioMC, carbotte, Phil, Ummarino, ilyapg}. Pad\'e approximants attempt to find the rational functions which interpolate the values of the electron or phonon Green's function or self-energy on the imaginary axis and capture the nonanalyticities in the lower half of the complex plane \cite{vidberg}. This method is not guaranteed to produce the correct analytic continuation especially at lower temperatures and does not always provide a faithful representation of the phonon structure at higher frequencies \cite{Ummarino, carbotte}. Attempts have also been made to use the extended Koopmans' theorem to perform analytic continuation of the imaginary axis Green's function but in practice the accuracy was comparable to that of Pad\'e approximants \cite{schuler}.

The GW approximation for perturbatively calculating the self-energy of a many-body system of electrons shares many similarities with the RME approximation as the self-energies are structurally the same. Within the GW community, there are several other methods which have been proposed for solving the real-axis GW equations which avoid the challenges of working directly on the real axis such as the contour deformation technique \cite{godby}. Other examples include plasmon-pole models for the screened Coulomb interaction or multipole models for the electron self-energy such as 2-pole models or Pad\'e approximants \cite{hybertsen, rojas, rieger, golze}. However, such methods often make restrictive assumptions which we would like to avoid and their accuracy is hard to judge.

Marsiglio et. al. provide an iterative method for analytic continuation of the UME equations in Ref. \onlinecite{Marsiglio} which avoids singular integrals and exhibits fast convergence. In this work, we extend this iterative method to the RME equations while maintaining the same benefits of the original method. The resulting analytic continuation is well-suited for numerical evaluation and converges quickly. Using this method, we obtain the electron and phonon spectral functions in the normal and superconducting states and take care to work in a regime of electron-phonon coupling strength where RME theory is valid. Treating both the electron and phonon propagators on an equal footing results in significant changes to the properties of the normal and superconducting states as compared to the case where the phonon propagator remains bare.  
 
ME theory provides a good description of the paradigmatic and widely-studied Holstein model in the adiabatic limit and the regime of weak-coupling \cite{Ilyame, MarsiglioMC, ilyapg, Alexandrov, berger, scalettar, costa, alder, ohgoe, freericks, ciuchi, freericks2, meyer, capone, hague}. The dimensionless electron-phonon coupling strength, $\lambda$, is an important parameter which generally controls whether the system exhibits metallic behavior, superconductivity, or charge-density wave order at low temperature \cite{Ilyame, ilyatc, ilyapg}. In this work we comment on differences in two common definitions of $\lambda$ which are accessible in experiments such as ARPES and electron tunneling \cite{cuk, mcmillan}. We find that $\lambda$ determined from electron mass renormalization is significantly different and less sensitive to superconducting correlations than $\lambda$ derived from the joint electron-phonon density of states at the Fermi level. We examine the analytically continued electronic spectral function in the superconducting state at the largest value of $\lambda$ for which RME theory is applicable to provide new insights on the maximal value of $2 \Delta / T_c$ attainable in conventional superconductors.

\section{\label{sec:level1}Models and methods}

We consider the Holstein Hamiltonian \cite{holstein} given by
\begin{equation}
\begin{split}
H &= \sum_{\kk,\sigma} \epsilon^{\phantom{\kk}}_{\kk} c_{\kk,\sigma}^\dagger c^{\phantom{\kk}}_{\kk,\sigma} + \Omega \sum_{\qq} \left( b^\dagger_{\qq} b^{\phantom{\qq}}_{\qq} + \frac{1}{2} \right)\\
&+\frac{\alpha}{\sqrt{N}} \sum_{\kk,\qq,\sigma} c_{\kk+\qq,\sigma}^\dagger c^{\phantom{\kk}}_{\kk,\sigma} \left( b^\dagger_{-\qq} + b_{\qq} \right)
\end{split}
\end{equation}
where $N$ is the number of lattice sites of a two-dimensional square lattice, $\Omega$ is the frequency of a dispersion-less Einstein phonon, $\epsilon_{\kk}$ is the band dispersion, $\alpha$ is a the electron-phonon coupling constant, $c^{\dagger}_{\kk,\sigma}$ creates an electron with momentum $\kk$ and spin $\sigma$, and we have set $\hbar=M=1$. In the case of a momentum independent el-ph coupling constant, we define the bare dimensionless electron-phonon coupling strength 
\begin{equation}
\lambda_0 = \frac{2 \alpha^2 \rho(E_F)}{\Omega}
\end{equation} 
where $\rho(E_F)$ is the density of states at the Fermi level.

In ME theory the equation for the electron self-energy on the imaginary axis is given by
\begin{align}
    \Sigma(\kk,i\omega_n) &= -\frac{\alpha^2}{\beta N} \sum_{\qq,m} D(\qq,i\nu_m) \tau_3 \nonumber \\
    &\quad\quad \times G(\kk-\qq,i\omega_n-i\nu_m) \tau_3
    \label{eq:imagsigma}
\end{align} 
and the phonon self-energy is given by
\begin{align}
    \Pi(\qq,i\nu_m) = \frac{\alpha^2}{\beta N} \sum_{\kk,n} \mathrm{Tr} &\Big[  G(\kk+\qq,i\omega_n+i\nu_m) \nonumber \\
    &\times\tau_3 G(\kk,i\omega_n) \tau_3 \Big]
    \label{eq:realsigma}
\end{align} 
where the $\tau_i$ are the Pauli matrices, $\beta$ is the inverse temperature, $\omega_n = 2n\pi/\beta$, $\nu_m=(2m+1)\pi/\beta$, $G(\kk, i\omega_n)=(i\omega_n \tau_0 - \epsilon_\kk \tau_3 - \Sigma(\kk, i\omega_n))^{-1}$, and $D(\qq, i\nu_m) = -2\Omega / \left(\nu_m^2 + \Omega^2 + 2 \Omega \Pi(\qq, i\nu_m) \right)$. $\Sigma(\kk, i\omega_n)$ and $G(\kk, i\omega_n)$ are $2 \times 2$ Nambu matrices with non-zero off-diagonal components in the superconducting phase. 

For numerical evaluation of Eq. (\ref{eq:imagsigma}, \ref{eq:realsigma}) we use the fast Fourier transform to perform convolutions in both imaginary frequency and momentum. We combine a higher-order evaluation of the Fourier integral described in Ref. \onlinecite{numericalrecipes} with a correction accounting for the discontinuity of the electronic Green's function in imaginary time. Linear mixing at every iteration with the result from the previous iteration is used to stabilize convergence of the self-energies during the self-consistent iterations.

The expressions for the retarded electron and phonon self-energies on the real axis can be obtained from the imaginary axis equations by introducing the spectral representation for the propagators, performing the Matsubara frequency summation, analytically continuing the result, and performing one of the contour integrals over frequency by making use of the analyticity of the retarded Green's function in the upper half-plane (for details of the derivation see appendix A). The resulting expressions are

\begin{widetext}
\begin{equation}
\begin{split}
\Sigma^R(\kk,\omega) &= -\frac{\alpha^2}{N}\sum_{\qq} \int_{-\infty}^\infty dz B(\qq,z) \tau_3 \left[ \frac{1}{\beta} \sum_m \frac{G(\kk-\qq,i\omega_m)}{\omega-i\omega_m-z} - G^R(\kk-\qq,\omega-z)\left[ 1 + n_B(z) - n_F(\omega-z)\right] \right] \tau_3 \\
\Pi^R(\qq,\omega) &= \frac{\alpha^2}{N}\sum_\kk \int_{-\infty}^\infty dz \mathrm{Tr} \left( A(\kk+\qq,z)\tau_3\left[ \frac{1}{\beta} \sum_{m} \frac{G(\kk,i\omega_m)}{\omega+i\omega_m-z} - G^A(\kk,z-\omega)\left[n_F(z-\omega)-n_F(z) \right] \right] \tau_3 \right)
\end{split}
\label{eq:real}
\end{equation}
\end{widetext}
where $G^{R/A}(\kk,\omega)$ is the retarded/advanced Green's function, $A(\kk,\omega) = -\frac{1}{\pi} \mathrm{Im} G^R(\kk,\omega)$ is the electronic spectral function, $B(\kk,\omega) = -\frac{1}{\pi} \mathrm{Im} D^R(\kk,\omega)$ is the phonon spectral function, $n_B(\omega)$ is the Bose-Einstein distribution function, and $n_F(\omega)$ is the Fermi-Dirac distribution function. The equation for the electronic self-energy $\Sigma^R(\kk,\omega)$ in Eq. \ref{eq:real} is derived in Ref. \onlinecite{Marsiglio}. The equation for the phonon self-energy $\Pi^R(\qq,\omega)$ is the additional equation needed to analytically continue RME theory. These equations reduce to the familiar non-selfconsistent single-iteration form for the electron and phonon self-energies by replacing the propagators with the bare expressions. In order to make use of these equations and to obtain the analytic continuation to the real axis, the ME equations are first solved self-consistently on the imaginary axis and the results are used as inputs to the pair of equations in Eq. \ref{eq:real} which are also solved self-consistently. In practice we find that the analytic continuation to the real axis converges quickly (typically requiring an order of magnitude fewer iterations than converging the imaginary axis calculation). The code is available at \href{https://github.com/bennosski/elph}{https://github.com/bennosski/elph}.

\begin{figure}
    \centering
    \includegraphics[width=8cm]{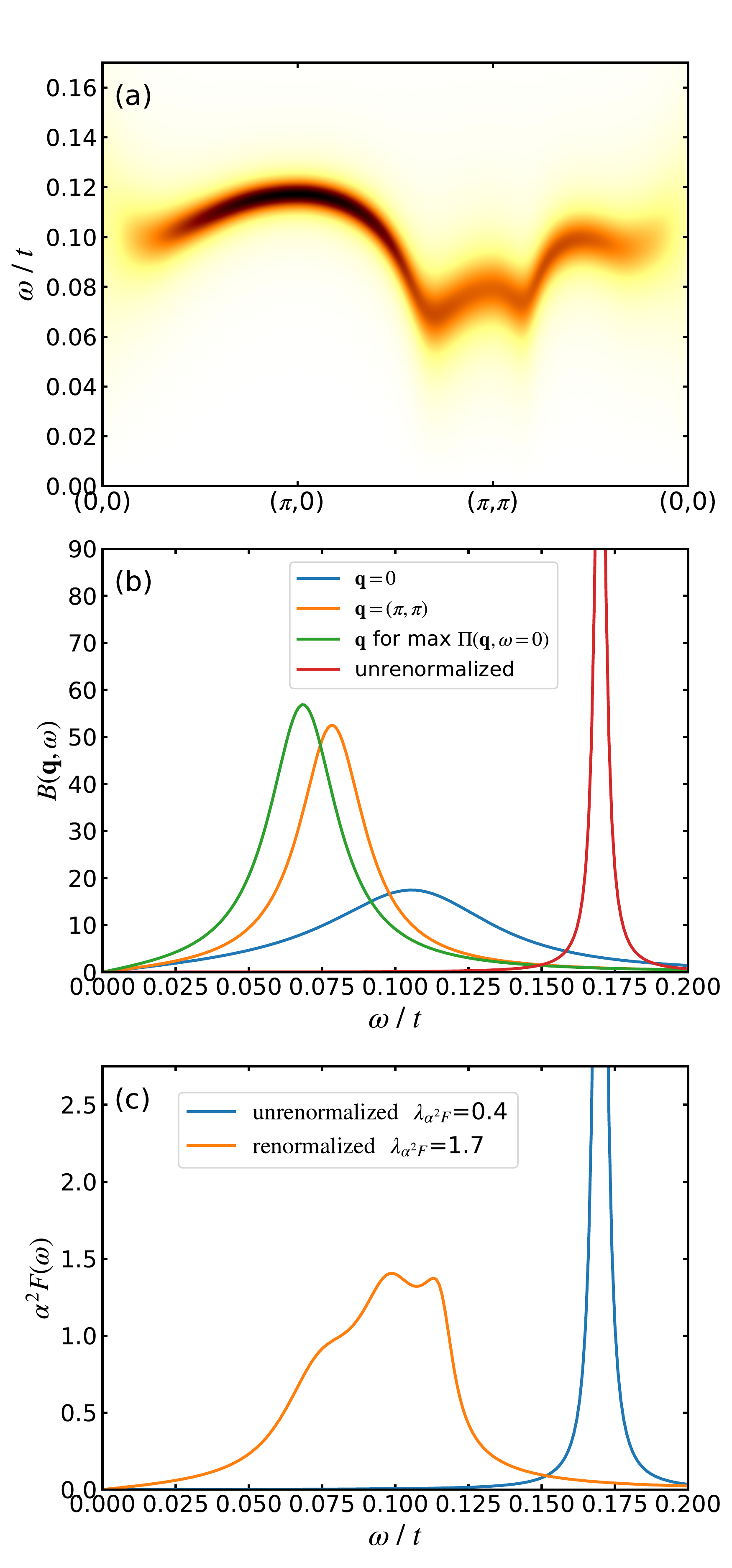}
    \caption{Holstein model in two dimensions on a $120 \times 120$ lattice, $t' = -0.3 t$, $n = 0.8$, $\lambda_0=0.4$, $\beta t = 16$, $\Omega / E_F = 0.1$. (a) Phonon spectral function for the RME theory. (b) Phonon spectral function at selected momentum points. (c) $\alpha^2 F(\omega)$ and corresponding $\lambda$ for UME theory (blue) and RME theory (orange).  }
    \label{fig:fig1}
\end{figure}

\section{\label{sec:level1} Results}

DQMC and other numerical studies have examined the accuracy of RME theory and have generally found it to be quantitatively accurate up to a critical value of $\lambda_0$ of order 1  \cite{Ilyame, Alexandrov, bauer}. Using the method of analytic continuation described above, we examine the real-axis spectral functions and self-energies obtained from RME theory within its range of validity. We choose parameters identical to as those in Ref. \onlinecite{Ilyame} representing a cuprate-like square-lattice bandstructure with next-nearest-neighbor hopping $t'=-0.3 t$ at an electron filling of $n = 0.8$ for which the density of states at the Fermi level for the bare band-structure is $\rho(E_F) t = 0.3$, an Einstein phonon with a phonon frequency of $\Omega = 0.17 t$ corresponding to an adiabatic ratio $\Omega/ E_F = 0.1$ for a dispersion-less Einstein mode, and a dimensionless el-ph coupling strength of $\lambda_0 = 0.4$ beyond which the ME theory rapidly breaks down and diverges from the DQMC results. For the parameter regime considered here, there is remarkably good quantitative agreement between ME theory and DQMC results for the single-particle electronic self-energy and superconducting susceptibility \cite{Ilyame}. In appendix B, we also show results for a half-filled band with $t'=0$ and at temperatures above the transition to the charge-density wave phase find the same qualitative results as for $n = 0.8$ and $t'=-0.3 t$.

\subsection{Normal state}

\begin{figure*}
    \centering
    \includegraphics[width=\textwidth]{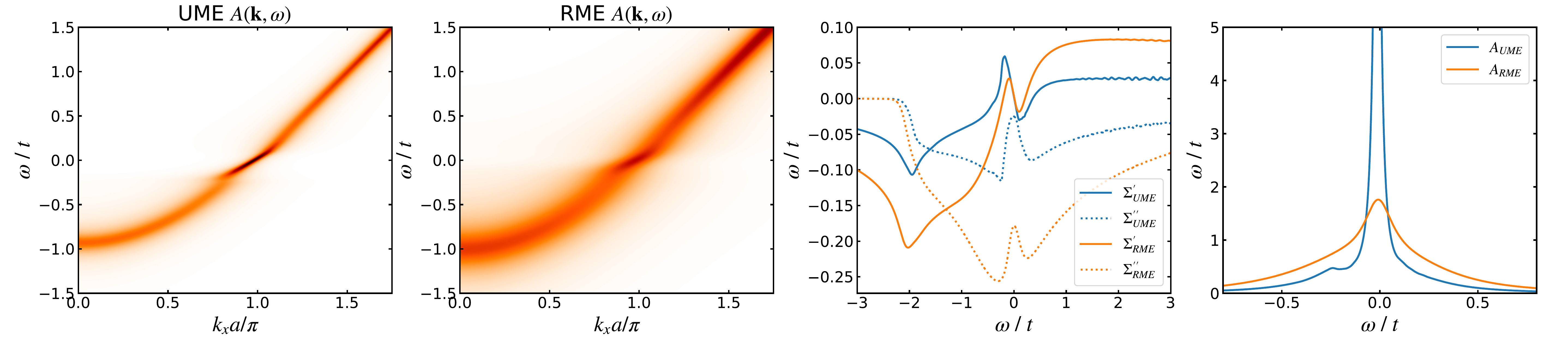}
    \caption{Holstein model in two dimensions on a $120 \times 120$ lattice, $t' = -0.3 t$, $n = 0.8$, $\lambda_0=0.4$, $\beta t = 16$, $\Omega / E_F = 0.1$. (a) Electronic spectral function for ME theory without renormalization with a momentum-space cut along ($0$, $\pi/2$) to ($\pi$, $\pi/2$). (b) Spectral function for RME theory for momentum-space cut along ($0$, $\pi/2$) to ($\pi$, $\pi/2$). (c) Real and imaginary parts of the electronic self-energies for UME theory (blue) and RME theory (orange) at the $\kk_{F}$ point along the momentum-space cut from ($0$, $\pi/2$) to ($\pi$, $\pi/2$). (d) Spectral functions for the renormalized and unrenormalized cases at the $\kk_{F}$ point along the momentum-space cut from ($0$, $\pi/2$) to ($\pi$, $\pi/2$). }
    \label{fig:fig2}
\end{figure*}

We first consider ME theory in the normal state at a temperature of $\beta t=16$ which is a temperature below the phonon energy scale. The solution of the self-consistent RME equations captures the renormalization of the phonon dispersion and linewidth due to electron-phonon interaction. Fig. \ref{fig:fig1} shows phonon-related quantities such as the phonon spectral function and the $\alpha^2 F(\omega)$ function introduced by McMillan and Rowell \cite{mcmillan}. The phonon frequency softens significantly from the bare $\Omega=0.17 t$ value as seen in Fig. \ref{fig:fig1}a especially at a wave-vector corresponding to the weak-coupling Fermi surface nesting for this band structure near $\qq = (\pi,\pi)$. The phonon spectral function for selected values of the wave-vector $\qq$ are shown in Fig. \ref{fig:fig1} illustrating the increased linewidth of the phonon mode caused by the el-ph interactions. The peak in the $a^2 F(\omega)$ function correspondingly moves towards lower frequency and broadens as shown in Fig. \ref{fig:fig1}c. The effective electron-phonon coupling is computed as $\lambda = 2 \int_0^\infty (\alpha^2 F(\omega) / \omega ) d\omega$ and is more than four times larger in the renormalized case ($\lambda=1.7$) as compared to the unrenormalized case ($\lambda=0.4$) due to a combination of increased spectral weight in the phonon spectral function due to an increase in the phonon occupation number and the shift of the peak of $\alpha^2F(\omega)$ to lower frequency. As will be discussed below, the larger effective electron-phonon coupling strength leads to an enhanced superconducting $T_c$ for the RME theory despite the softening of the phonon. We note that these observations are similar to the behavior observed in RME theory outside the adiabatic limit with $\Omega=1t$ as noted in Ref. \onlinecite{Phil} which is a limit where RME theory appears to be less reliable based on comparison with DQMC \cite{Ilyame}.

The electronic spectra for the RME theory exhibit a significant broadening/incoherence compared to the unrenormalized case. As seen in the imaginary part of of the electronic self-energy in Fig. \ref{fig:fig2}b, the broadening is over $2.5$ times larger in the renormalized case which can be understood as a result of the increased number of phonons in the renormalized calculation caused by the tendency towards lattice instability as signified by the softening of the phonon. Based on eq. \ref{eq:real}, the imaginary part of the electronic self-energy has a term proportional to the the phonon spectral function times the Bose occupation factor convolved with the electronic density states, a quantity which increases as the phonon softens since $\int d\omega n_B(\omega) B(\qq,\omega) = 2N_q+1$ where $N_q$ is the number of phonons at wave-vector $\qq$. In the UME calculation the average number of phonons per site is $\langle N \rangle = \frac{1}{2}\left[n_B(\Omega) - n_B(-\Omega) - 1 \right] = 0.07$ and in the RME calculation we find $\langle N \rangle = 0.85$. The qualitative difference in the spectra can be seen in the spectral functions shown in Fig. \ref{fig:fig2}d at a momentum point on the Fermi surface. Note that the good agreement between RME theory and DQMC suggests that this intrinsic broadening would not be undone by including higher order diagrams or vertex corrections. It is also interesting to note that the real part of the electronic self-energy is similar between UME and RME theory near the Fermi energy despite the significant differences in the imaginary part. As will be discussed later, this means the electron mass renormalization is similar between the renormalized and unrenormalized cases despite much stronger effective el-ph coupling. The results for a half-filled system without next-nearest-neighbor hopping ($t'=0$) also in the $\Omega/E_F < 1$ limit and below the charge-density wave ordering temperature are qualitatively similar to the results described here and can be found in appendix B. Parameters were chosen to match those of Ref. \onlinecite{MarsiglioMC} where it was also shown that the superconducting and charge-density wave susceptibilities from RME theory agree better with DQMC calculations than those from UME theory.

\subsection{Superconductivity}

\begin{figure}
    \centering
    \includegraphics[scale=0.45]{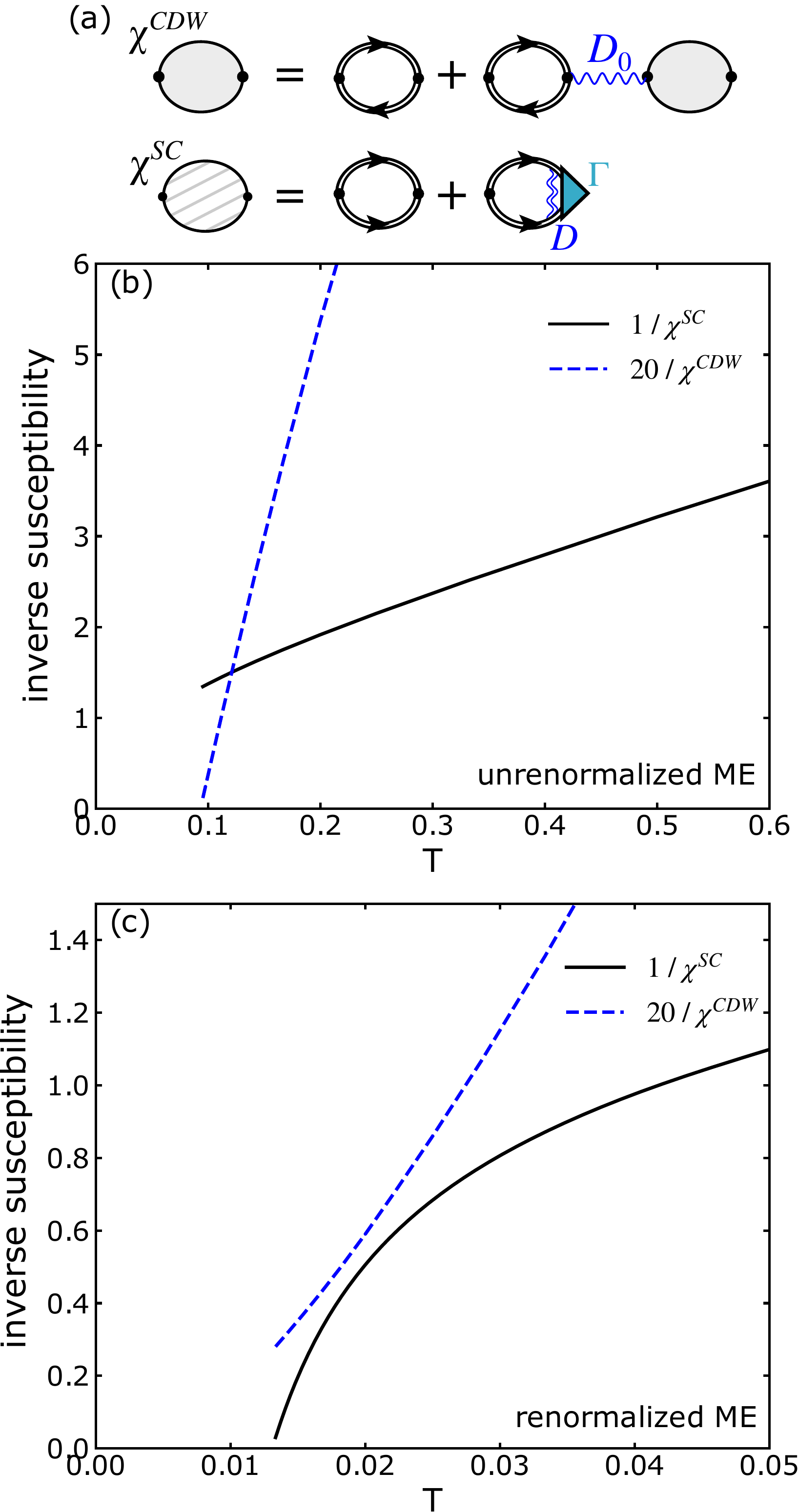}
    \caption{Inverse of the superconducting and charge-density wave susceptibilities for the Holstein model in two dimensions on a $128 \times 128$ lattice, $t' = -0.3 t$, $n = 0.8$, $\lambda_0=0.4$, and $\Omega = 0.17t$ (corresponding to $\Omega / E_F = 0.1$). (a) Diagrams for the charge-density wave and superconducting susceptibilities in the Migdal approximation. (b) UME theory. (c) RME theory. }
    \label{fig:susceptibilities}
\end{figure}

Including the phonon self-energy in the ME theory in fact changes the ground state of the system from a charge-density wave to a superconductor in the weak-coupling, adiabatic regime considered here. This can be seen from the temperature dependence of the charge-density-wave susceptibility, $\chi^{CDW}$, and the superconducting susceptibility, $\chi^{SC}$, in the normal state. We compute the susceptibilities within the Migdal approximation illustrated by the diagrams in Fig. \ref{fig:susceptibilities}a which correspond to summing the series of particle-hole ring diagrams for the charge-density wave susceptibility and summing the series of particle-particle ladder diagrams for superconducting susceptibility. The charge-density wave susceptibility is given by

\begin{equation}
\begin{split}
\chi^{CDW}(\qq) &= \frac{\chi_0(\qq,i\nu_0)}{1 + \alpha^2 D_0(\qq,i\nu_0)\chi_0(\qq,i\nu_0)} \\
\chi_0(\qq,i\nu_m) &= -\frac{2}{N \beta} \sum_{\kk,n} G(\kk,\iwn) G(\kk+\qq,\iwn+i\nu_m) \\
\end{split}
\end{equation}
and the superconducting susceptibility is given by
%\begin{widetext}
\begin{equation}
\begin{split}
\chi^{SC}(\qq) &= \frac{1}{N \beta} \sum_{\kk,m} F(\kk,\iwm) \Gamma(\kk,\iwm) \\
F(\kk,\iwm) &= G(\kk,\iwm)G(-\kk,-\iwm) \\
\Gamma(\kk,\iwn) &= 1 - \frac{\alpha^2}{N \beta} \sum_{\qq,m} F(\kk-\qq,\iwn-\ivn) D(\qq, i\nu_m) \\
&\ \ \ \ \ \ \ \ \ \ \ \ \ \ \ \ \ \ \ \ \Gamma(\kk-\qq,\iwn-i\nu_m).
\end{split}
\end{equation}
%\end{widetext}

The charge-density-wave susceptibility is suppressed and the superconducting susceptibility is enhanced for RME theory relative to UME theory as seen in Fig. \ref{fig:susceptibilities}b which plots the inverse of the susceptibilities versus temperature. In the unrenormalized case, the charge-density-wave susceptibility diverges before the superconducting susceptibility indicating the system enters the charge-density-wave phase at a temperature of around $0.1t$. In the renormalized case, the charge-density-wave susceptibility is relatively large in magnitude but does not appear to tend toward a divergence at a finite temperature. The superconducting susceptibility narrowly wins and diverges at a finite temperature of $T_c = 0.013t$. The enhancement of superconductivity in the renormalized case make sense given the larger effective el-ph coupling strength.

\begin{figure}
    \centering
    \includegraphics[scale=0.45]{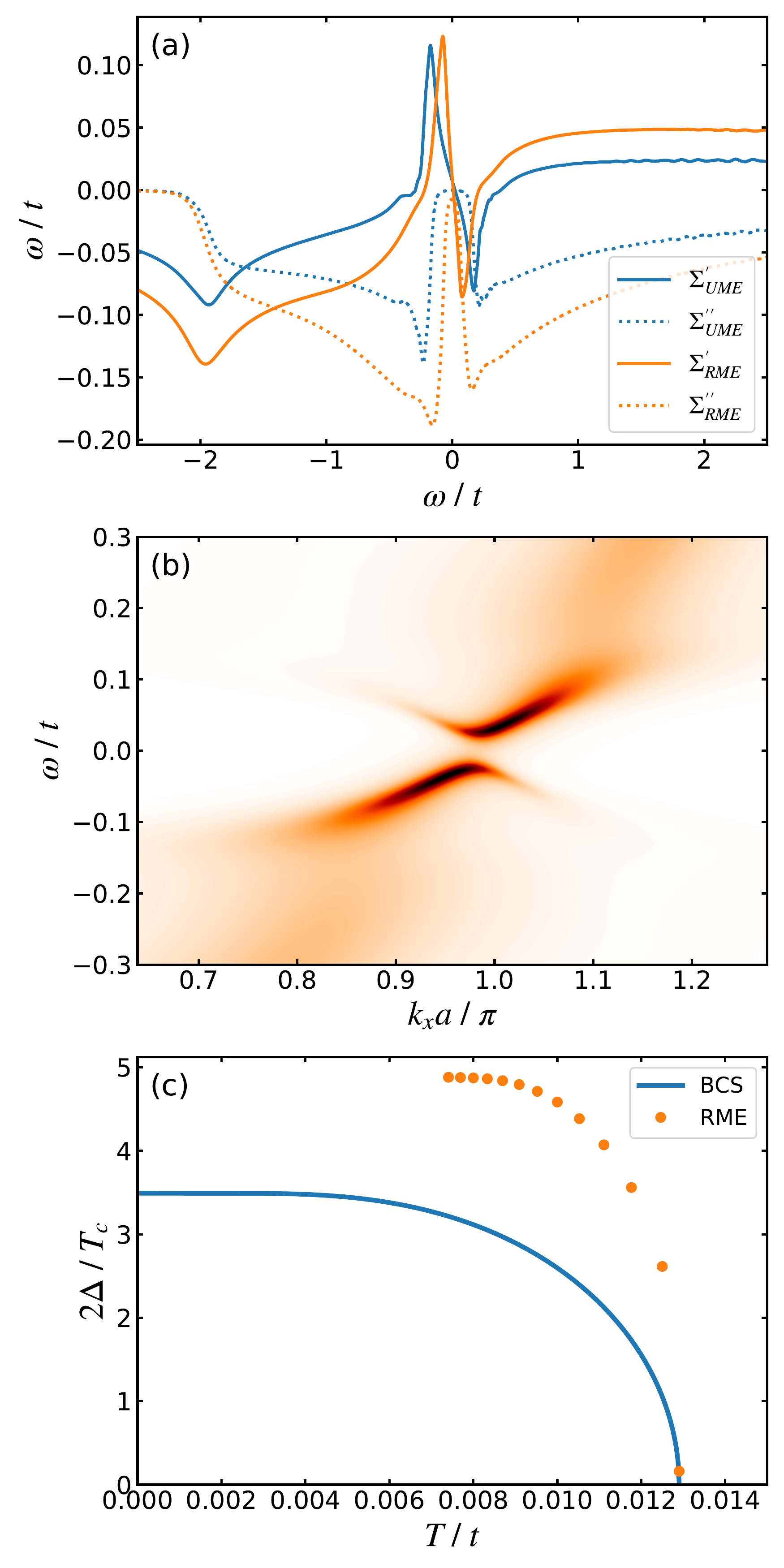}
    \caption{Holstein model in two dimensions on a $128 \times 128$ lattice, $t' = -0.3 t$, $n = 0.8$, $\lambda_0=0.4$, $\beta t = 100$, and $\Omega = 0.17t$ (corresponding to $\Omega / E_F = 0.1$). (a) Real (solid) and imaginary (dotted) parts of the diagonal components of the electronic self-energies for UME theory (blue) and RME theory (orange) at the antinode. (b) Electronic spectral function for RME theory for momentum-space cut along ($0$, $\pi/2$) to ($\pi$, $\pi/2$). (c) Superconducting order parameter (gap size) as a function of temperature for RME theory. Blue line is the gap function obtained from BCS theory. }
    \label{fig:fig3}
\end{figure}

We next consider the same system at a lower temperature of $\beta t = 100$ in order to access the superconducting phase. As shown in Fig. \ref{fig:fig3}, the imaginary part of the magnitude of electronic self-energy for the renormalized calculation is less significantly different than the imaginary part of the self-energy for the unrenormalized calculation because the total number of phonons decreases at low temperatures. The $0.03 t$ superconducting gap is visible in the electronic spectral function in Fig. \ref{fig:fig3}b. The superconducting order parameter, given by the size of the superconducting gap, is calculated as $\Delta(\kk) = \frac{\Sigma_{12}(\kk,\omega)}{Z(\omega)} |_{\omega=0}$ where $\omega(1 - Z(\kk,\omega)) = \frac{1}{2}\left[ \Sigma_{11}(\kk,\omega) - \Sigma_{11}(\kk,-\omega) \right]$ and the subscript indices on the self-energy indicate either the diagonal or off-diagonal Nambu components. The superconducting transition temperature based on the temperature at which the superconducting order parameter goes to zero is $T_c = 0.013 t$, consistent with the transition temperature obtained from the divergence of the superconductivity susceptibilities. 

Within the very weak coupling limit of BCS theory, the ratio of twice the superconducting gap to the transition temperature is set by a universal value given by $2 \Delta / T_c = 2\pi/e^\gamma = 3.53$. For strongly coupled superconductors $2 \Delta / T_c$ generally exceeds the BCS value \cite{mitrovic, carbotte}. Within ME theory, $2 \Delta / T_c$ closely follows a phenomenological form which increases monotonically as a function of the ratio set by $T_c$ to the phonon energy scale \cite{mitrovic}. It has recently been argued that the ratio of $T_c$ to the phonon frequency for conventional superconductors is bounded by $T_c/\Omega < 0.1$ because superconductivity is suppressed at larger values of $\lambda_0$ by strong-coupling physics such as the formation of polarons and/or CDW order. With these considerations in mind, $2\Delta / T_c$ should be maximized in a conventional el-ph mediated superconductor when $T_c / \Omega \approx 0.1$ as any further increase in $\lambda_0$ will decrease $T_c$ as well as $2\Delta/T_c$. 

In our calculation, $T_c/\Omega = 0.08$ which is close to saturating the upper bound for conventional superconductors and, as shown in Fig. \ref{fig:fig3}c, the order parameter as a function of temperature does not fit to the BCS form and results in $2\Delta/T_c \approx 5$. We therefore expect that the $2 \Delta / T_c \approx 5$ ratio represents an approximate upper bound for conventional superconductors with pairing mediated by a strongly retarded electron-phonon interaction. Indeed, the $2 \Delta / T_c$ ratios experimentally observed in many conventional superconductors as illustrated in Ref. \onlinecite{mitrovic} are generally within the range $2 \Delta / T_c \lesssim 5$. We do not expect that tuning the details of the electronic band-structure or shape of the Fermi surface would affect these results significantly because the band-structure considered here avoids any special Fermi surface nesting conditions. Choosing a Fermi surface favoring a particular wave-vector for nesting would increase CDW correlations and suppress superconductivity and the $2 \Delta / T_c$ ratio.

\subsection{Strength of el-ph coupling}

Quantifying the strength of electron-phonon coupling in various materials is of general interest and importance, especially considering the fact that the optimal regime for superconductivity in conventional superconductors is sensitive to the electron-phonon coupling strength and occurs at intermediate electron-phonon coupling strength. Up to this point we have discussed the strength of electron-phonon coupling in terms of the dimensionless electron-phonon coupling constant obtained from integrating the $\alpha^2 F(\omega)$ function which is experimentally accessible by electron tunneling experiments under the assumption that ME theory provides a good description of the system with an electron-boson spectral function $\alpha^2 F(\omega)$ describing the effective electron-electron interactions due to a generic boson exchange and the existence of a well-defined Fermi surface \cite{mcmillan, carbotte, jansen}. However, another common definition of electron-phonon coupling strength is based on the mass enhancement of electronic quasiparticles near the Fermi surface which can be measured directly through experimental techniques such as ARPES \cite{cuk}. The coupling constant derived from the mass renormalization at a given point on the Fermi surface can be defined in terms of the real part of the electronic self-energy as \cite{mahan}
\begin{equation}
\lambda_m(\kk_F) = - \frac{\partial \Sigma^\prime(\kk_F,\omega)}{\partial \omega} |_{\omega=0}.
\end{equation}
The phonon density of states is of course also k-dependent and the $\alpha^2 F(\omega)$ function previously used to calculate $\lambda_{\alpha^2 F}$ is in fact calculated as a Fermi surface average for which the equivalent k-dependent $\lambda$ for the Holstein model is
\begin{equation}
\lambda_{\alpha^2 F}(\kk_F) = \frac{2 \alpha^2}{(2 \pi)^2} \oint \frac{d \kk}{v_F(\kk)} \int_0^\infty \frac{d\omega}{\omega} B(\kk_F - \kk, \omega)
\end{equation}
where $\oint$ denotes integration over the Fermi surface, $v_F(\kk)$ is the Fermi velocity. The Fermi surface average performed to compute $\lambda_m$ from $\lambda_m(\kk_F)$ and $\lambda_{\alpha^2 F}$ from $\lambda_{\alpha^2 F}(\kk_F)$ is given by
\begin{equation}
\lambda = \frac{\oint \frac{d\kk}{v_F(\kk)} \lambda(\kk)}{ \oint \frac{d\kk}{v_F(\kk)} }.
\end{equation}

\begin{figure}
    \centering
    \includegraphics[width=8cm]{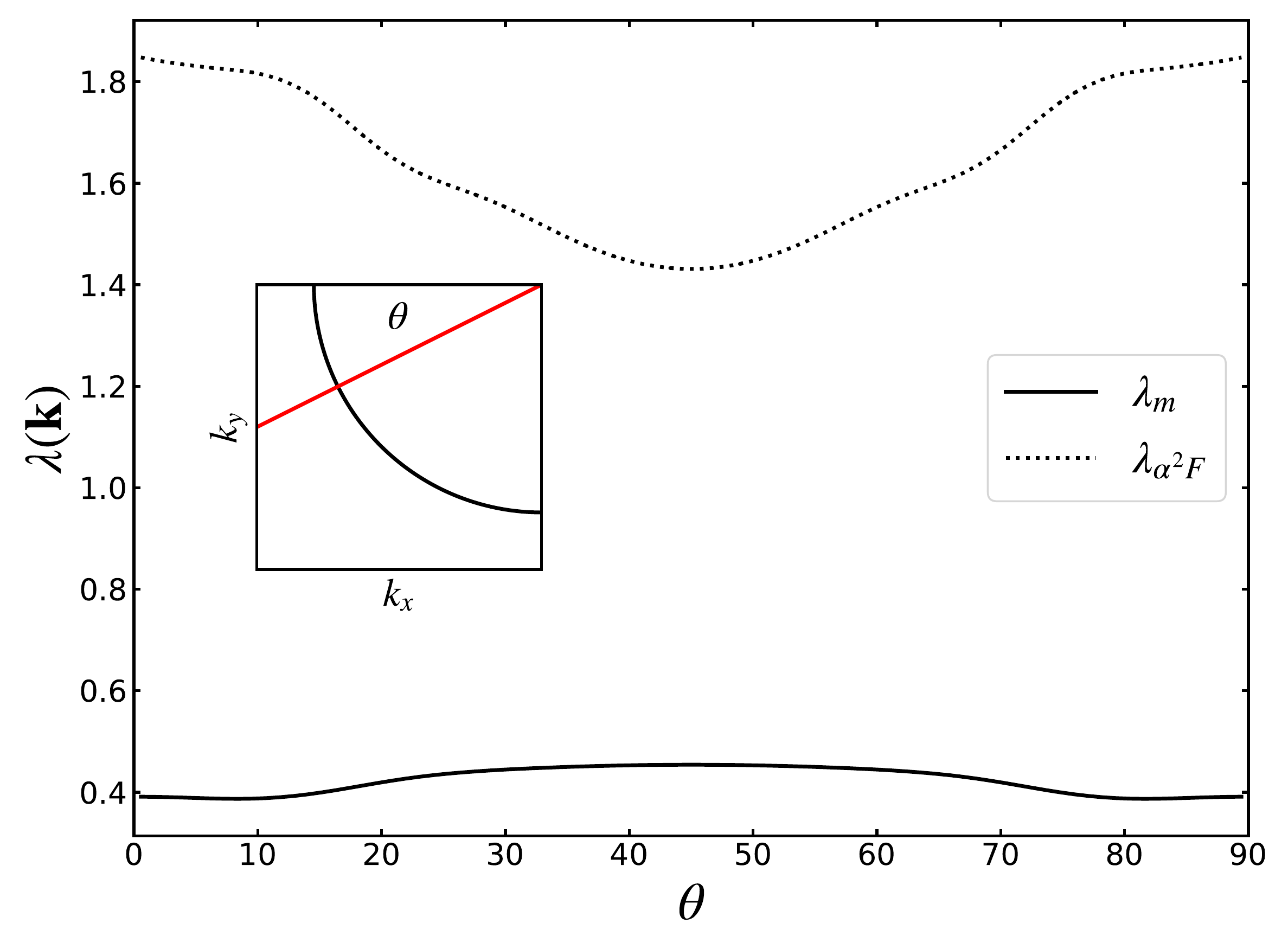}
    \caption{Fermionic momentum dependence of the el-ph coupling along the Fermi surface calculated for RME theory for the Holstein model in two dimensions on a $120 \times 120$ lattice, $t' = -0.3 t$, $n = 0.8$, $\beta t = 16$,  $\lambda_0=0.4$, and $\Omega = 0.17t$ (corresponding to $\Omega / E_F = 0.1$). Inset shows the Fermi surface and the definition of the angle $\theta$. }
    \label{fig:lambkr}
\end{figure}

Clearly $\lambda_m$ and $\lambda_{\alpha^2 F}$ are two different definitions of the dimensionless electron-phonon coupling. The definition of $\lambda_m$ is not directly sensitive to the phonon density of states or the lifetime/broadening of the electronic quasiparticles. $\lambda_m$ and $\lambda_{\alpha^2 F}$ are in principle related assuming a constant density of states and low temperature \cite{mahan}. However, for the general solution of the ME theory, $\lambda_m$ and $\lambda_{\alpha^2 F}$ are not equivalent, especially for the case with phonon renormalization. For the normal state calculation considered above, we find that $\lambda_m = 0.4$ and $\lambda_{\alpha^2 F} = 0.4$ for UME theory, and $\lambda_m = 0.4$ and $\lambda_{\alpha^2 F} = 1.7$ for RME theory. The similar values of $\lambda_m$ are consistent with the similar slopes seen in the real part of the electronic self-energy near zero frequency as shown in Fig. \ref{fig:fig2}c. Within RME theory, the momentum dependence of $\lambda_m$ and $\lambda_{\alpha^2 F}$ is even qualitatively different such that $\lambda_m$ is maximized at the node while $\lambda_{\alpha^2 F}$ is maximized at the antinode as shown in Fig. \ref{fig:lambkr}.

The Holstein model is known to exhibit competition between SC and CDW/(bi)polaron phases \cite{MarsiglioMC, ilyapg, Alexandrov, berger, scalettar, costa, alder, ohgoe, freericks, ciuchi, freericks2, meyer, capone, hague}. The tendency towards formation of polarons is associated with an increase of $\lambda_m$ as the electrons become heavier due to being dressed by a cloud of phonons. Since the value of $\lambda_m$ observed for both UME and RME theory in the normal state is the same, the tendency towards formation of CDW order or polarons does not change significantly with phonon renormalization in the regime studied here. The same conclusion can be drawn based on the similarity in magnitude of $\lambda_0$ and $\lambda_m$. 

In contrast to the behavior of $\lambda_m$, we can associate the significant increase of $\lambda_{\alpha^2 F}$ with the enhancement of superconductivity observed in the renormalized calculation. Therefore it appears $\lambda_{\alpha^2 F}$ is likely more informative and sensitive to  superconducting correlations than $\lambda_m$. In summary, by including phonon self-energy, the superconducting tendency is enhanced more significantly than the polaronic/CDW tendency based on an analysis of $\lambda_m$ versus $\lambda_{\alpha^2 F}$ as well as the strengths of the superconducting and charge-density wave susceptibilities. Given the significant (factor of four) difference between $\lambda_m$ and $\lambda_{\alpha^2 F}$ in our calculations, care should be taken when drawing conclusions about the effective strength of el-ph coupling as is relevant to superconductivity or relating $\lambda_m$ to $\lambda_{\alpha^2 F}$.

Although it is evident from the BCS expression $T_c \approx \omega_D \exp(-1/\lambda)$ (and similarly by the $\omega_{\mathrm{ln}}$ prefactor in the formula for $T_c$ obtained for ME theory by McMillan, Allen and Dynes \cite{mcmillantc, allendynes}) that a larger (bare) phonon frequency is favorable for superconductivity, the softening of a phonon mode can be counteracted by a boost in the electron-phonon coupling strength arising from increased spectral weight in the phonon spectral function (increased phonon occupation number) and a shift of the phonon density of states to lower frequency. The same behavior is seen away from the adiabatic limit with $\Omega/t = 1$ in Ref. \onlinecite{Phil}. Therefore softening of a phonon mode does not necessarily suppress superconductivity. This is surprising based on the intuition that phonon softening is generally associated with stronger charge correlations which are antagonistic towards superconductivity. However it appears that at least in the regime studied here, phonon softening is not always directly indicative of the strength of charge correlations as is evident by the suppression of the charge-density wave susceptibility and the minimal change of electron effective mass. Including the effects of phonon self-energy in the calculations can actually tip the scales in the delicate balance of CDW/SC competition in favor of superconductivity.

\section{Conclusion}

RME theory accounts for the interaction between electrons and phonons at the same lowest-order diagrammatic level of approximation for both the electron and phonon self-energies and, in contrast to UME theory, quantitatively agrees with numerically exact results from DQMC in the weak-coupling, adiabatic limit. The method described in this work allows for a numerically efficient analytic continuation of the solution of the imaginary-axis solutions of the RME equations, providing an improvement over approximate methods of analytic continuation such as Pad\'e approximants and can be applied to a range of self-energy approximations such as the GW approximation \cite{golze}. The real-frequency solution of the RME equations can provide insight into the renormalization of the phonon propagator in weakly coupled electron-phonon systems which is accessible to experiments such as inelastic x-ray scattering and neutron scattering. Renormalization of the phonon propagator can enhance the effective electron-phonon coupling strength ($\lambda_{\alpha^2 F}$) without significant electron mass enhancement ($\lambda_m$), resulting in an enhancement of superconductivity as compared to calculations performed with unrenormalized phonons. The dimensionless electron-phonon coupling strength derived from electron mass renormalized and the McMillan function $\alpha^2 F(\omega)$ are quantitatively and qualitatively different when significant renormalization of the phonon mode occurs. The resulting superconducting state exhibits a $2 \Delta / T_c$ ratio larger than that of the very weak-coupling limit described by BCS theory. As our calculation resides in the adiabatic limit at a critical value of electron-phonon coupling beyond which superconductivity is suppressed by strong-coupling physics, the $2 \Delta / T_c \approx 5$ ratio sets an approximate upper bound for conventional superconductors with phonon-mediated pairing with strongly retarded interactions.

\begin{acknowledgments}
We thank Ilya Esterlis, Edwin Huang, Yoni Schattner, Brian Moritz, Philip Dee, Zhi-Xun Shen, Steven Johnston, Steven Kivelson and Frank Marsiglio for illuminating discussions. This work was supported by the U.S. Department of Energy (DOE), Office of Basic Energy Sciences, Division of Materials Sciences and Engineering under Contract no. AC02-76SF00515. Computational work was performed on the Sherlock cluster at Stanford University supported. M. S. thanks the Alexander von Humboldt Foundation for its support with a Feodor Lynen scholarship.
\end{acknowledgments}

% \onecolumngrid
\appendix

\section{Derivation of real-axis equations}
This section presents the derivation of the equation for the real-axis phonon self-energy in Eq. \ref{eq:real}. The derivation of the electronic self-energy is available in Ref. \onlinecite{Marsiglio}. The phonon self-energy on the imaginary axis is given by
\begin{equation}
\Pi^R(\qq, \ivn) = \frac{2 \alpha^2}{N \beta} \sum_{\kk,m} G(\kk, \iwm) G(\kk+\qq,\iwm+\ivn) \ .
\end{equation}
Introducing the spectral representation for the Green's function $G(\iwm, \kk) = \int_{-\infty}^\infty dz \frac{A(\qq,z)}{i\omega_m-z}$ and performing the Matsubara frequency sum yields

\begin{widetext}

\begin{equation}
\Pi^R(\qq,\ivn) = \frac{2 \alpha^2}{N} \sum_{\kk,m} \int_{-\infty}^\infty dz dz' \frac{A(\kk,z) A(\kk+\qq, z')\left[n_F(z)-n_F(z')\right]}{\ivn+z-z'}
\end{equation}
\\
\\
Performing the analytic continuation $i\nu_n \rightarrow \omega+i\delta$ and using $A(\kk,z) = -\frac{1}{\pi}\imG$ gives
\begin{equation}
\Pi^R(\qq,\omega+i\delta) = -\frac{2 \alpha^2}{N \pi} \sum_{\kk,m} \int_{-\infty}^\infty dz dz' \frac{ A(\kk+\qq,z')\imG \left[n_F(z)-n_F(z')\right]}{\omega+z-z'+i\delta}
\label{eq:a1}
\end{equation}
The final form of the real-axis equation which avoids principal value integrations is obtained by analytically performing the integral over $z$ in Eq. \ref{eq:a1}. To perform this integral consider the following integral of only those parts of Eq. \ref{eq:a1} which depend on $z$.

\begin{equation}
\mathrm{I} = \int_{-\infty}^\infty dz dz' \frac{ \imG \left[n_F(z)-n_F(z')\right]}{\omega+z-z'+i\delta}
\end{equation}
The next step is to separate this integral into separate integrals for the real and imaginary parts in order to make use of the analyticity of $G^R(z,\kk)$ in the upper half-plane for contour integration. Defining $\omega_{\pm} = \omega - z - z' \pm i\delta$, the integral becomes

\begin{align}
I  &= \mathrm{Im} \left\lbrace \int_{-\infty}^\infty dz \left[n_F(z)-n_F(z')\right] \frac{G^R(\kk,z)}{2}\left(\frac{1}{\omega_+} + \frac{1}{\omega_-} \right) \right\rbrace \nonumber \\ 
& \quad - i \mathrm{Re} \left\lbrace \int_{-\infty}^\infty dz \left[n_F(z)-n_F(z')\right] \frac{G^R(\kk,z)}{2} \left( \frac{1}{\omega_+} - \frac{1}{\omega_-} \right) \right\rbrace
\end{align}
These integrals can be evaluated by considering a contour integral over the upper complex plane for which the relevant poles are $z = \omega - z' + i\delta$ and $z = i (2m + 1) \pi/ \beta$ for $m \geq 0$.
% \begin{equation}
% \begin{split}
% \int_{-\infty}^\infty dz \left[n_F(z) &- n_F(z')\right] \frac{G^R(\kk,z)}{2}\left(\frac{1}{\omega_+} \pm \frac{1}{\omega_-} \right) = a \pm b \\    
% a &= - \frac{\pi i}{\beta}\sum_{m=0}^{\infty} G(\kk,\iwm) \frac{1}{\omega+\iwm-z'+i\delta} \\
% b &= \pi i \left[n_F(z'-\omega+i\delta) - n_F(z')\right] G^R(\kk,z'-\omega+i\delta) - \frac{\pi i}{\beta}\sum_{m=0}^{\infty} G(\kk,\iwm) \frac{1}{\omega+\iwm-z'-i\delta} 
% \end{split}
% \end{equation}
Let us define the consider
\begin{align*}
	% \int_{-\i´nfty}^\infty dz \left[n_F(z) &- n_F(z')\right] 
	\int_{-\infty}^\infty dz [n_F(z)- n_F(z')]\frac{G^R(\kk,z)}{2}\left(\frac{1}{\omega_+} \pm \frac{1}{\omega_-} \right) \equiv a \pm b 
	 \ ,
\end{align*}
with
\begin{align*}
	 a &= - \frac{\pi i}{\beta}\sum_{m=0}^{\infty} G(\kk,\iwm) \frac{1}{\omega+\iwm-z'+i\delta}  \ , \\
	 b &= \pi i \left[n_F(z'-\omega+i\delta) - n_F(z')\right] G^R(\kk,z'-\omega+i\delta) - \frac{\pi i}{\beta}\sum_{m=0}^{\infty} G(\kk,\iwm) \frac{1}{\omega+\iwm-z'-i\delta}  \ .
\end{align*}

The integral $I$ is then given by

\begin{equation}
\begin{split}
I &= \mathrm{Im} \left( a + b \right) - i \mathrm{Re} \left( a - b \right) 
 = -ia + ib^* \\
&= \pi \left[n_F(z'-\omega) - n_F(z')\right]G^{A}(z'-\omega,\kk) -\frac{\pi}{\beta}\sum_{m=-\infty}^{\infty} \frac{G(\kk,\iwm)}{\omega+\iwm-z'}
\end{split}
\end{equation}
where $G^A(\kk,\omega) = [G^R(\kk,\omega)]^*$, the relation $G^*(\kk,\iwm) = G(\kk,-\iwm)$ was used, and $i\delta$ was dropped from the denominator of the second term since it is small relative to $i\omega_m$. Returning to Eq. \ref{eq:a1} gives the final result

\begin{equation}
\Pi^R(\qq,\omega+i\delta) = \frac{2\alpha^2}{N} \sum_\kk \int_{-\infty}^\infty dz A(\kk+\qq,z)\left[ \frac{1}{\beta} \sum_{\kk,m}  \frac{G(\kk,\iwm)}{\omega+\iwm-z} - G^{A}(\kk,z-\omega)\left[n_F(z-\omega) - n_F(z)\right] \right]
\end{equation}
One can check that by inserting the from of the non-interacting $G(\kk,\iwm) = (\iwm - \epsilon_\kk)^{-1}$ and $G^R(\kk,\omega+i\delta) = (\omega-\epsilon_\kk + i\delta)^{-1}$ into the final result yields the correct form for the usual single-iteration phonon self-energy given by

\begin{equation}
\Pi^R_0(\qq,\omega+i\delta) = \frac{2\alpha^2}{N} \sum_\kk \frac{n_F(\epsilon_\kk) - n_F(\epsilon_{\kk+\qq}) }{\omega+i\delta+\epsilon_\kk-\epsilon_{\kk+\qq}}
\end{equation}.

\begin{figure*}
    \centering
    \includegraphics[width=\textwidth]{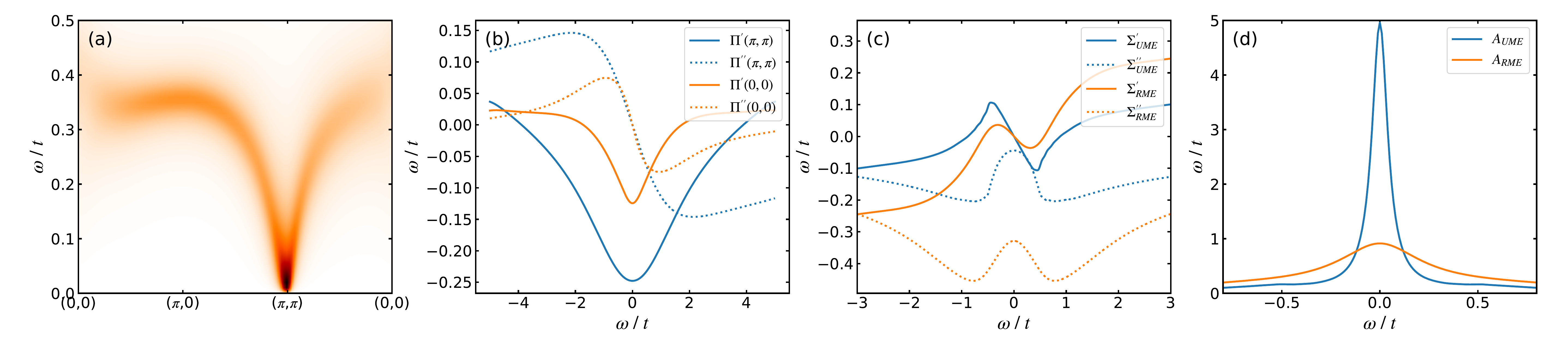}
    \caption{Holstein model in two dimensions on a $80 \times 80$ lattice, $n = 1$, $t' = 0$, $\alpha^2 / \Omega^2 = 1.5 t$, $\beta t = 6$, $\Omega = 0.5t$. (a) Phonon spectral function. (b) Real and imaginary parts of the phonon self-energy for RME theory at $\kk=(\pi,\pi)$ and $\kk=(0,0)$. (c) Real and imaginary parts of the electronic self-energies for UME theory (blue) and RME theory (orange) at $\kk_F = (\pi/2, \pi/2)$. (d) Electronic spectral functions for the renormalized and unrenormalized cases at $\kk_F = (\pi/2, \pi/2)$. }
    \label{fig:figb}
\end{figure*}

\end{widetext}

\section{Holstein model at half filling}

We consider the case of a two dimensional Holstein model at half-filling and without next-nearest-neighbor hopping ($t'=0$). Parameters were chosen to match those in Ref. \onlinecite{MarsiglioMC} which demonstrates that RME theory agrees with DQMC results whereas the UME theory does not for these parameters. The phonon frequency is $\Omega=0.5t$ corresponding to $\Omega/E_F = 0.25$. The electron phonon coupling strength is set by $\alpha^2 / \Omega^2 = 1.5 t$. The spectral functions and self-energies are qualitatively similar to those for the parameters in the main text. In the half-filled case the softening of the phonon occurs exactly at the $\qq=(\pi,\pi)$ wave-vector and is already very strong at the temperature of $\beta t = 6$ due to the strong charge-density wave instability from perfect Fermi surface nesting. The major difference between the renormalized and UME theory is the significantly broader electronic spectra in the renormalized case as can be seen in Fig. \ref{fig:figb} (d).

%\nocite{*}


\end{document}